\documentclass[twocolumn,showpacs,preprintnumbers,amsmath,amssymb, floatfix]{revtex4}

\usepackage{graphicx}
\usepackage{dcolumn}
\usepackage{bm}

\newcommand{\id}{{\mathbf 1}}

\newcommand{\rmi}{{\mathrm i}}
\newcommand{\rme}[1]{{\mathrm e}^{#1}}

\newcommand{\R}{{\mathbb R}}

\newcommand{\sign}{\mathrm{sign}}

\begin{document}

\preprint{APS/123-QED}

\title{Visualizing the kinematics of relativistic wave packets}

\author{Bernd Thaller}
\affiliation{%
Institute for Mathematics and Scientific Computation, University of Graz\\
Heinrichstrasse 36, A-8010 Graz, Austria
}%

\date{\today}

\begin{abstract}
This article investigates some solutions of the time-dependent free Dirac equation.
Visualizations of these solutions immediately reveal strange phenomena 
which are caused by the interference of positive- and negative-energy waves.
The effects discussed here include the {\textsl{Zitterbewegung}}, the opposite direction of momentum and velocity in negative-energy wave packets, and the superluminal propagation of the wave packet's local maxima.
\end{abstract}


\maketitle

\section{Introduction}

The numerical simulation of the Schr\"odinger equation and the visualization of its solutions has become an important part of quantum-mechanical education on all levels \cite{B: Brandt,B: Feagin,B: Horbatsch,VQM}.
But, it  is almost unavoidable to stumble accross strange phenomena, when one attempts a numerical solution of the Dirac equation.
In this article we show visualizations of some of these peculiarities and try to give an informal  explanation.

The Dirac equation is the fundamental equation for relativistic quantum mechanics.
Hence, it belongs to the most important equations in modern physics.
Among its big successes is the very accurate description of the energy levels of the hydrogen atom.
On the other hand, the occurrence of several unexplained paradoxes casts doubts on its status and interpretation.

The unexpected behavior occurs even for innocent looking initial conditions.
A canonical set of initial conditions for the time-dependent Schr\"odinger or Dirac equations is given by the set of Gaussian wave packets.
They describe more or less localized quantum states for which the product of the uncertainties in position and momentum is minimal.
On the other hand, the set of Gaussian initial conditions covers most cases of practical interest, because any wave packet can be approximated by a superposition of a finite number of Gaussian states.

\begin{figure}
\includegraphics[width=2.4in]{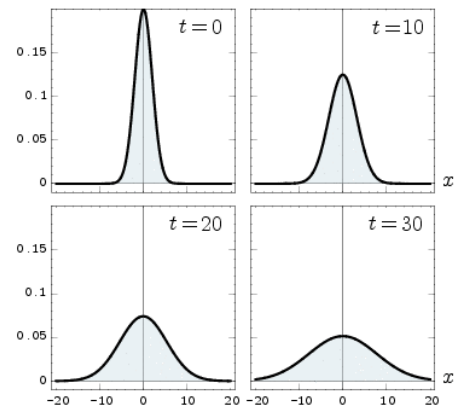}
\caption{\label{fig:g} Spreading of a Gaussian wave packet according to the Schr\"odinger equation.}
\end{figure}

The motion of Gaussian wave packets according to the one-dimensional free Schr\"odinger equation
shows little surprises. The example in Fig.~\ref{fig:g} shows a nonrelativistic Gaussian wave packet with average momentum zero. Initially, the wave packet is well localized, but it spreads during the time evolution.
As the wave packet gets smeared out, its height decreases, because its norm as a square-integrable function must remain constant.
The wave packet remains a Gaussian function for all times.

\begin{figure}
\includegraphics[width=2.4in]{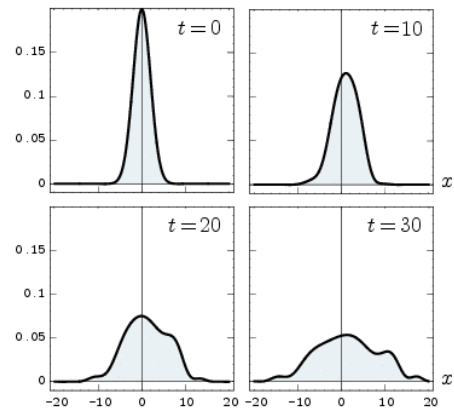}
\caption{\label{fig:a} Time evolution of a Gaussian initial wave packet according to the Dirac equation.}
\end{figure}

According to the free Schr\"odinger equation, the average position $\langle{\mathbf x}\rangle$ and the average momentum $\langle{\mathbf p}\rangle$ of the wave packet obey the rules of classical mechanics.
Moreover, the spreading of the wave packet is independent of the average velocity of the wave packet.
This spreading of the position distribution would also be observed for a cloud of classical particles whose density in position space is a Gaussian function, provided that the momenta of the particles also have a Gaussian distribution.

Fig.~\ref{fig:a} shows a numerical solution of the one-dimensional free Dirac equation.
It has the same Gaussian initial distribution as the nonrelativistic wave packet in Fig.~\ref{fig:g}, yet its behavior is quite different.
The relativistic wave packet wiggles back and forth, becomes non-Gaussian for $t\ne 0$, and soon develops characteristic ripples. 

This result is so strange that anybody with some experience with quantum mechanics (but not with the Dirac equation) would first assume that the numerical method is at fault.
This is a good example supporting the argument that the numerical solution of an equation is rarely sufficient to understand a phenomenon.
This motivates a more careful theoretical analysis in order to understand the origin of this strange behavior. 

This article contains several black-and-white images of solutions of the Dirac equation.
Computer-generated animations showing the whole time evolution can be found on the internet \cite{URL}.
These movies show both components of the solution with a color code for the phase of complex numbers.
Thus they reveal more information than the simple black-and-white reproductions of the position probability density in this article.
The forthcoming book \cite{AVQM} will be accompanied by a CD-ROM containing a large collection of similar movies, animations, and simulations. They illustrate and supplement the theoretical exposition in order to provide students with an intuitive understanding that is hard to achieve by studying mathematical formulas.
See the web site \cite{VQMURL} for more details and the ideas behind the project Visual Quantum Mechanics.

\section{The Dirac equation}

In this article, we discuss the time-dependent free Dirac equation in one space dimension.
We write it as an evolution equation in the familiar ``Schr\"odinger form''
\begin{equation}\label{free Dirac equation}
{\rmi} \,\hbar \, \frac{d}{dt}\,\psi(x,t) = H_0\,\psi(x,t), \qquad \psi(x,0) = \psi_0(x).
\end{equation}
The free Dirac Hamiltonian $H_0$ is the matrix-differential operator
\begin{equation}\label{free Dirac operator}
H_0 =  c \,\sigma_1 \,p + \sigma_3 \,mc^2,
\end{equation}
where $\sigma_1$ and $\sigma_3$ are the famous Pauli matrices, 
and $p=-\rmi\,\hbar\,d/dx$.

The expression for $H_0$ can be interpreted as a linearization of the relativistic energy-momentum relation
\begin{equation}
E = \lambda(p) = \sqrt{c^2 p^2 + m^2 c^4}.
\end{equation}
The square of the Dirac operator is just given by
\begin{equation}
H_0^2 = c^2 p^2 + m^2 c^4.
\end{equation}

For numerical computations and for the visualizations it is advantageous to use units where $\hbar = m = c = 1$. These can be obtained from the SI-units by a simple scaling transformation. Hence, in the following we use Dirac equation in the form
\begin{equation}\label{free Dirac equation dimless}
{\mathrm i}  \, \frac{\partial}{\partial t}\,\psi(x,t) = \Bigl(-\rmi\,\sigma_1\,\frac{\partial}{\partial x} + \sigma_3\Bigr)\,\psi(x,t).
\end{equation}
Instead of $\sigma_1$ and $\sigma_3$, we could use any other pair of Pauli matrices.
This would give a unitarily equivalent formulation.
All images in this article would remain unchanged.

The phenomena explained here also occur in higher dimensions, but the one-dimensional situation is much easier to visualize.
For the Dirac Hamiltonian in three space-dimensions, Pauli matrices are not sufficient; $4\times 4$-Dirac matrices are needed instead.
The book \cite{Thaller92} contains more information about the Dirac equation in three dimensions.

\section{Dirac spinors and their interpretation}

A suitable state-space for the solutions of the Dirac equation must consist of vector-valued functions
\begin{equation}\label{E: two components}
\psi(x, t) = \begin{pmatrix} \psi_1(x, t)\\ \psi_2(x, t)\end{pmatrix},
\end{equation}
because the operator $H_0$ is a two-by-two matrix.
These two-component wave functions are usually called Dirac spinors.
The word ``spinor'' might appear inappropriate.
In one dimension, all magnetic fields that could possibly affect the spin are pure gauge fields. 
Hence, the two components of \eqref{E: two components} do not describe the spin.
Rather, they reflect the appearance of negative energies. (However, the doubling of the spinor components in three dimensions is caused by the spin.)

The free Dirac Hamiltonian in momentum space is the matrix
\begin{equation}\label{E: Dirac operator}
{\mathrm h}_0(p) = \begin{pmatrix} mc^2 & cp\\ cp & - mc^2\end{pmatrix},
\end{equation}
which for each $p\in\R$ has the eigenvalues $\lambda (p)$ and $-\lambda(p)$.
As the Hamiltonian represents the energy of the quantum system, one is led to the conclusion  that the possible values of the energy of a free particle can be positive or negative.

Correspondingly, the Dirac equation has two types of plane-wave solutions, which we denote by $u_{\mathrm{pos}}$ and $u_{\mathrm{neg}}$.
For each $p\in\R$,
\begin{equation}\label{E: plane waves time dep pos}
u_{\genfrac{}{}{0pt}{}{\mathrm{pos}}{\mathrm{neg}}}(p;x,t) = \frac{1}{\sqrt{2\pi}}\,u_{\genfrac{}{}{0pt}{}{\mathrm{pos}}{\mathrm{neg}}}(p)\,\rme{\rmi p x \mp \rmi \lambda(p) t},
\end{equation}
where $u_{\mathrm{pos}}(p)$ and $u_{\mathrm{neg}}(p)$ are eigenvectors of the matrix ${\mathrm h}_0(p)$ belonging to the eigenvalues $\lambda(p)$ and $-\lambda(p)$, respectively.
Hence, we have
\begin{equation}\label{E: dirac eigenfunctions}
H_0 \,u_{\genfrac{}{}{0pt}{}{\mathrm{pos}}{\mathrm{neg}}}(p;x,t)= \pm \lambda(p)\,u_{\genfrac{}{}{0pt}{}{\mathrm{pos}}{\mathrm{neg}}}(p;x,t)
\end{equation}
and therefore $u_{\mathrm{pos}}$ ($u_{\mathrm{neg}}$) is called a plane wave with positive (negative) energy.

In order to be in line with the formalism of quantum mechanics, one usually requires
that (for each $t$) the components of a Dirac-spinor be square-integrable, 
\begin{equation}\label{E: spinor square integrability}
\int_{-\infty}^\infty |\psi_j(x,t)|^2\,dx < \infty \quad\text{(for all $t$ and $j=1,2$).}
\end{equation}
One should be aware that this mathematical requirement
is closely related to the interpretation of the solutions. Our choice makes perfect sense if we
agree with the following tentative interpretation of relativistic wave functions:

Suppose $\psi$ is a normalized Dirac spinor, then
\begin{equation}\label{E: tent interpretation pos}
\int_{a}^b \bigl(  |\psi_1(x,t)|^2 + |\psi_2(x,t)|^2\bigr)\,dx 
\end{equation}
is the probability to find the particle in the interval $(a.b)\subset \R$.

Here ``normalized'' means that the above integral from $-\infty$ to $+\infty$ is equal to $1$.
Similarly, and in complete analogy to nonrelativistic quantum mechanics,
\begin{equation}\label{E: tent interpretation mom}
\int_{a}^b \bigl(  |\hat\psi_1(p,t)|^2 + |\hat\psi_2(p,t)|^2\bigr)\,dp 
\end{equation}
is interpreted as the probability to find the momentum of the particle in the interval from $a$ to $b$.
(The hat denotes the Fourier transform with respect to $x$.)
This interpretation is consistent with the choice of $p=-\rmi d/dx$ (the generator of spatial translations) as the momentum operator.

The unitarity of the time-evolution generated by the Dirac equation guarantees that
the normalization of Dirac spinors is time-independent.
Hence, if there is one particle at the beginning, there will be one particle at the end.
This remains true for arbitrary external fields.
Contrary to a common belief, the Dirac equation is not able to describe pair production within the framework of quantum mechanics.
For this one needs the formalism of quantum field theory where many-particle wave functions are built
from one-particle solutions (and from one-antiparticle solutions, which are related to negative-energy solutions).

Square-integrable wave packets can be obtained from the plane waves by superposition.
This is the same procedure that works for the Schr\"odinger equation. With suitable coefficient functions, any square-integrable solution of the Dirac equation can be written in the form
\begin{align}
\psi(x,t) = {} \int_{-\infty}^\infty \bigl( & \hat\psi_{\mathrm{pos}}(p)\,u_{\mathrm{pos}}(p;x,t)
\nonumber
\\
\label{E: wave packets time ev}
& + \hat\psi_{\mathrm{neg}}(p)\,u_{\mathrm{neg}}(p;x,t)\bigr)\,dp .
\end{align}
The coefficient functions $\hat\psi_{\mathrm{pos}}$ and $\hat\psi_{\mathrm{neg}}$ can be determined from the Fourier transform $\hat\psi(p,0)$ of the initial function $\psi(x,0)$ by a projection onto the positive or negative energy subspace:
\begin{equation}\label{E: projection pos en}
 \hat\psi_{\genfrac{}{}{0pt}{}{\mathrm{pos}}{\mathrm{neg}}}(p) = P_{\genfrac{}{}{0pt}{}{\mathrm{pos}}{\mathrm{neg}}}\,\hat\psi(p,0) = \frac 12\Bigl(\id \pm \frac{{\mathrm h}_0(p)}{\lambda(p)}\Bigr)\,\hat\psi(p,0).
 \end{equation}
Numerical integration of \eqref{E: wave packets time ev} provides a method to compute the free time evolution of an arbitrary initial function that can be a useful alternative to a finite difference scheme.
More about the mathematics of the one-dimensional Dirac equation can be found in \cite{AVQM}.

For many reasons, including some of the observations in this article, the interpretation given above
may be critisized.
Therefore, one should consider this interpretation merely as a convenient working hypothesis
that allows us to apply the well-established mathematical formalism of quantum mechanics.

\section{Examples of relativistic kinematics}

In this section, we describe and visualize three strange phenomena shown by solutions of the free Dirac equation. Some explanations will be given in the following sections.

As a first example, we compute the free time evolution of the Gaussian Dirac spinor
\begin{equation}\label{E: gaussian wave packet 1}
\psi(x,0) = \Bigl(\frac 1{32 \pi}\Bigr)^{1/4}\exp(-x^2/16)\,\begin{pmatrix} 1\\ 1 \end{pmatrix}.
\end{equation}
A few snapshots of the solution are shown in Fig.~\ref{fig:a}.
This image shows the position probability density $ |\psi_1(t,x)|^2 + |\psi_2(t,x)|^2$ according to the interpretation given above. Compared to the simple spreading of the nonrelativistic Gaussian, this wave packet shows a very complicated motion.

\begin{figure}
\includegraphics[width=2.4in]{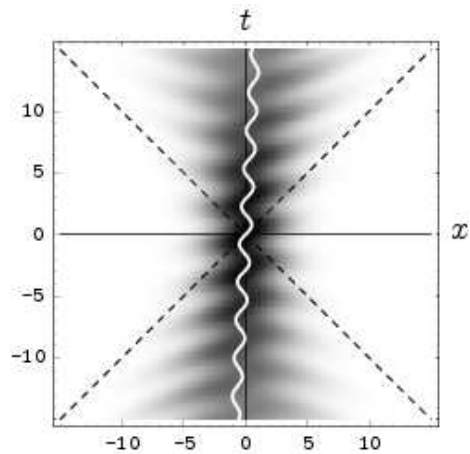}
\caption{\label{fig:aworld} Space-time diagram and worldline of the position mean value for the wave packet in Fig.~\ref{fig:a}.}
\end{figure}

The choice of the numerical constants in \eqref{E: gaussian wave packet 1} is motivated by the following consideration.
The Fourier transform of the wave packet \eqref{E: gaussian wave packet 1} is
\begin{equation}\label{E: gaussian wave packet 1mom}
\hat\psi(p,0) = \Bigl(\frac{2}{\pi}\Bigr)^{1/4}\exp(-4p^2)\,\begin{pmatrix} 1\\ 1 \end{pmatrix}.
\end{equation}
It describes a momentum distribution that is well localized in the interval $[-3/4,+3/4]$.
This corresponds to a maximal speed less than $3/5$, far below the speed of light $c=1$.
Hence, we can be sure that all observations have little to do with relativistic effects due to velocities approaching the speed of light.

Fig.~\ref{fig:aworld} shows a space-time diagram of this solution.
Here, the density plot visualizes the position probability density as a function of the space coordinate $x$ and the time coordinate $t$ in dimensionless units.
The white curve shows the world line of the average position, which obviously does not obey classical (relativistic) kinematics. Instead, the expectation value of the position operator performs a wiggling motion, commonly known as ``Zitterbewegung'' \cite{Schr32}.
Apart from the rapid oscillation, the wave packet drifts slowly to the right, although its average momentum is zero.
Nevertheless, it turns out that the momentum distribution (as determined from the Fourier transform of the wave packet) is still a conserved quantity with average momentum zero.

\begin{figure}
\includegraphics[width=2.4in]{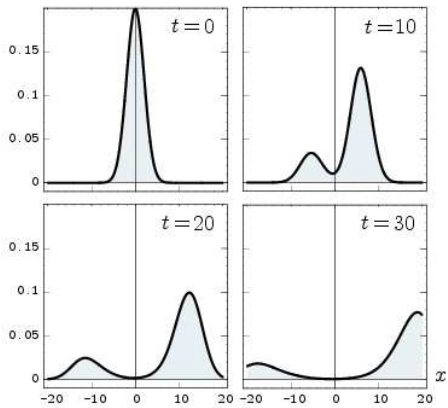}
\caption{\label{fig:b} Time evolution of a Gaussian initial wave packet with positive momentum.}
\end{figure}

The second example, Fig.~\ref{fig:b} shows the free time evolution of the initial spinor
\begin{equation}\label{E: gaussian wave packet 3}
\psi(x,0) = \Bigl(\frac 1{32 \pi}\Bigr)^{1/4}\exp(-x^2/16 - \rmi 3x/4)\,\begin{pmatrix} 1\\ 1 \end{pmatrix}.
\end{equation}

This spinor is very similar to \eqref{E: gaussian wave packet 1}. In position space, it is multiplied by a phase factor $\exp(-\rmi 3 x/4)$. In momentum space, this means a translation by $3/4$.
The momentum distribution belonging to \eqref{E: gaussian wave packet 3} is therefore a Gaussian distribution centered around the average momentum $3/4$.
Moreover, the momentum distribution is so narrow that all momenta contributing significantly to the wave packet are positive.
Nevertheless, the solution splits into two parts, and the smaller part moves to the left (that is, with a negative velocity).

\begin{figure}
\includegraphics[width=2.4in]{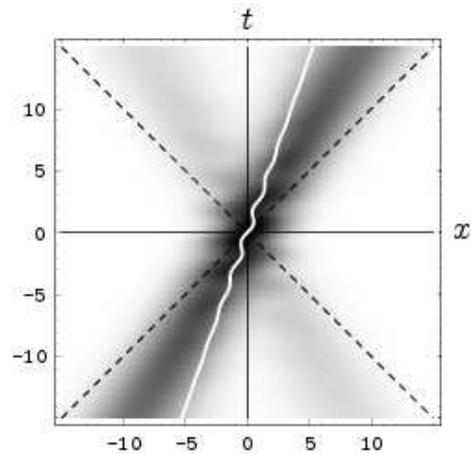}
\caption{\label{fig:bworld} Space-time diagram and worldline of the position mean value for the wave packet in Fig.~\ref{fig:b}.}
\end{figure}

The space-time diagram of this solution, shown in Fig.~\ref{fig:bworld}, again shows the Zitterbewegung of the position's mean value. But this time, the oscillation quickly fades away. We see that Zitterbewegung is sustained only as long as the left-moving part and the right-moving part of the wave packet have some overlap in position space.

\begin{figure}
\includegraphics[width=2.4in]{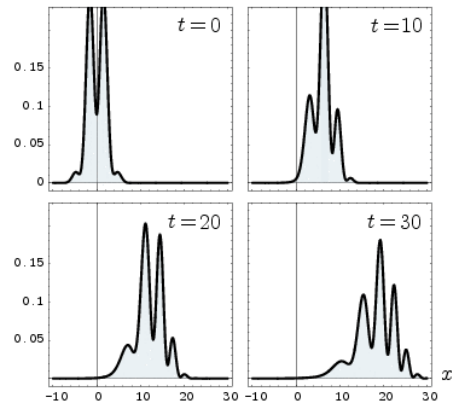}
\caption{\label{fig:c} Time evolution of a wave packet with positive velocity.}
\end{figure}

The third example is shown in Fig.~\ref{fig:c}. It realizes a wave packet with positive velocities. The initial wave packet was obtained as a superposition of a positive- and a negative-energy part:
\begin{equation}\label{E: gaussian wave packet 4}
\psi(x,0) = \psi_{\mathrm{pos}}(x) +  \psi_{\mathrm{neg}}(x),
\end{equation}
The positive-energy part has positive momentum. In momentum space, it is defined as
\begin{equation}\label{E: gaussian wave packet 4a}
\hat\psi_{\mathrm{pos}}(x) = N P_{\mathrm{pos}}\,\exp\bigl(-4(p-4/5)^2\bigr)\,\begin{pmatrix} 1\\ 0 \end{pmatrix}.
\end{equation}
The negative-energy part has negative momentum,
\begin{equation}\label{E: gaussian wave packet 4b}
\hat\psi_{\mathrm{neg}}(x) = N P_{\mathrm{neg}}\,\exp\bigl(-4(p+4/5)^2\bigr)\,\begin{pmatrix} 0\\ 1 \end{pmatrix}
\end{equation}
In position space, both parts obviously move in the same direction. They interfere with each other, thereby causing the ripples in the position distribution.

\begin{figure}
\includegraphics[width=2.4in]{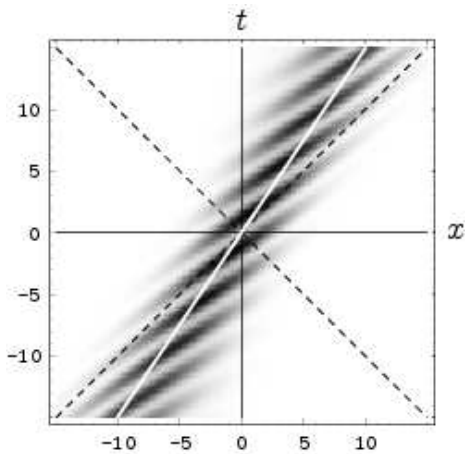}
\caption{\label{fig:cworld} Space-time diagram and worldline of the position mean value for the wave packet in Fig.~\ref{fig:c}.}
\end{figure}

Fig.~\ref{fig:cworld} shows a space-time diagram of this motion. There is no Zitterbewegung at all; the world line of the average position is a straight line. Interestingly, the peaks of the wave packet move with superluminal speed. This is strange because, at least in principle, local variations of the position probability density are observable. Let us now proceed with a theoretical analysis.

\section{Parity and direction of motion}

The average velocity of the wave packet \eqref{E: gaussian wave packet 1} shows a slow drift to the right, which is clearly visible in Fig.~\ref{fig:aworld}.
It is remarkable that the solution is not symmetric with respect to reflections at the origin, although the initial condition is. Indeed, the Dirac equation is not invariant under the replacement
\begin{equation}\label{E: space reflection}
\psi(x,t) \to \psi(-x,t).
\end{equation}
But the replacement $x\to -x$ in the wave function does not describe the physical space reflection. Like any Lorentz transformation, the space reflection has a part that acts on the components of the wave function. The correct way to describe the space reflection in the Hilbert space of the Dirac equation is the parity transform
\begin{equation}\label{E: parity transform}
P:\psi(x,t) \to \sigma_3\psi(-x,t).
\end{equation}
Hence, the wave packet \eqref{E: gaussian wave packet 1} is not invariant under a parity transform, $P\psi$ is not a scalar multiple of $\psi$.

\begin{figure}
\includegraphics[width=2.4in]{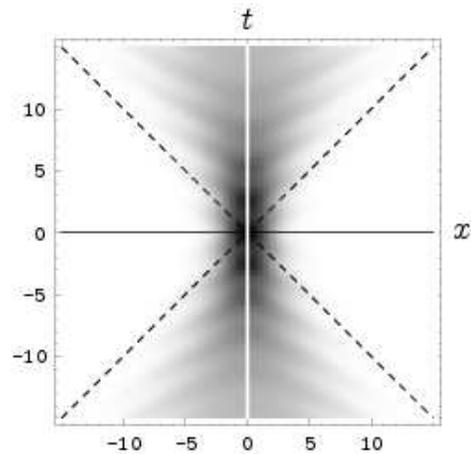}
\caption{\label{fig:dworld} Space-time diagram and worldline of the position mean value for the wave packet \eqref{E: gaussian wave packet 2}.}
\end{figure}

An example of a parity invariant solution is provided by the Gaussian initial wave packet
\begin{equation}\label{E: gaussian wave packet 2}
\psi(x,0) = \Bigl(\frac 1{4\pi}\Bigr)^{1/4}\exp(- x^2/8)\,\begin{pmatrix} 1\\ 0 \end{pmatrix},
\end{equation}
which has only an upper component and satisfies $P\psi(x,t) = \psi(x,t)$ for all times.
Fig.~\ref{fig:dworld} shows a space-time diagram of the position probability density for this solution.
The average velocity is zero, and the wave packet shows no Zitterbewegung in the expectation value---although the position probability density shows ripples similar to the first solution.

The average velocity of a wave packet is described by the classical velocity operator
\begin{equation}\label{E: classical velocity operator}
v_{\mathrm{cl}}(p) =  c^2\,p\,H_0^{-1} .
\end{equation}
It corresponds to the definition of the velocity in terms of momentum and energy, which is familiar from classical relativistic mechanics.
Note that this relation between velocity and momentum depends, in particular, on the sign of the energy.
For a wave packet with negative energy, a positive momentum $p$ thus corresponds to a negative average velocity $v_{\mathrm{cl}}$.
This is also reflected by the phase velocity of the plane waves \eqref{E: plane waves time dep pos}. The phase velocity of the plane waves is
\begin{equation}\label{E:phase velocity}
v_{\mathrm{ph}} = \sign(E)\,\frac{\lambda(p)}{p}.
\end{equation}
Note that the phase velocity is always faster than the velocity of light, and even tends to infinity in the limit of small momenta.
This does not matter, because no information can be transmitted with phase velocity.
Actually, a plane wave is spread over all of space-time and all information that is carried by a plane wave is already everywhere.
The group velocity of relativistic wave packets is always slower than or equal to the velocity of light.
But the sign of the phase velocity carries over to the sign of the group velocity.

\begin{figure}
\includegraphics[width=1.8in]{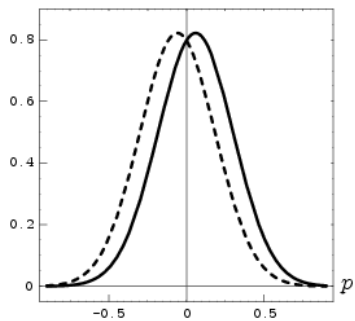}
\caption{\label{fig:amom} Momentum distributions of the positive and negative energy parts of the wave packet in Fig.~\ref{fig:amom}.}
\end{figure}

For the wave packet \eqref{E: gaussian wave packet 1}, Fig.~\ref{fig:amom} shows the momentum distributions of the parts with positive and negative energy, that is, the functions $|\hat\psi_{\mathrm{pos}}(p)|^2$ and $|\hat\psi_{\mathrm{neg}}(p)|^2$, according to \eqref{E: projection pos en}.
We see that the positive-energy part has its momentum distribution slightly shifted towards positive momenta, whereas the negative-energy part has a negative average momentum.
For a positive-energy wave packet, a positive average momentum means a positive average velocity, as usual.
But, for a negative-energy wave packet, the negative average momentum corresponds to a positive average velocity.
Hence, the whole wave packet has a positive average velocity, which can be seen clearly in Fig.~\ref{fig:aworld}.

\begin{figure}
\includegraphics[width=1.8in]{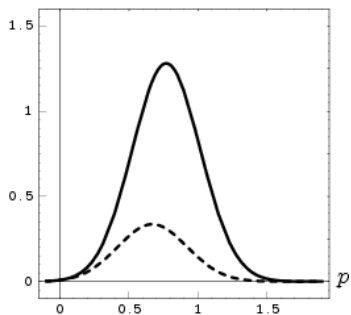}
\caption{\label{fig:bmom} Momentum distributions of the positive and negative energy parts of the wave packet in Fig.~\ref{fig:b}.}
\end{figure}

The opposite direction of velocity and momentum in the negative-energy part of a wave packet immediately explains the behavior of the wave packet \eqref{E: gaussian wave packet 3} in Fig.~\ref{fig:b}. 
For this wave packet, the momentum distributions of the parts with positive and negative energy are shown in Fig.~\ref{fig:bmom}. We see that both parts  consist of positive momenta. For the smaller negative-energy part, this corresponds to a negative velocity. Hence, the wave packet in position space is a superposition of an approximately Gaussian wave packet with positive energy moving to the right, and a smaller part with negative energy moving to the left (also with a positive average momentum).

\begin{figure}
\includegraphics[width=2.1in]{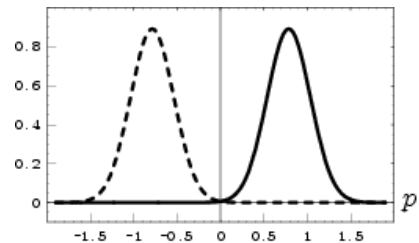}
\caption{\label{fig:cmom} Momentum distributions of the positive and negative energy parts of the wave packet in Fig.~\ref{fig:c}.}
\end{figure}

In case of the third example \eqref{E: gaussian wave packet 4}, the negative-energy part has negative momenta, as shown in Fig.~\ref{fig:cmom}.
Here the two parts are of the same size and they have opposite momenta, they have the same average velocity.
Hence, the parts with positive and negative energy move together in position space. The opposite phase-directions of the two parts cause the interference effects.
At the average momentum $\langle p\rangle$, the initial state contains the superposition of the plane waves $u_{\mathrm{pos}}(\langle p\rangle;x,0) + u_{\mathrm{neg}}(-\langle p\rangle;x,0)$.
In both components of the spinor, this is proportional to a superposition of $\exp(\rmi \langle p\rangle x)$ and $\exp(-\rmi \langle p\rangle x)$.
This, precisely, causes the sine-like distortions of the Gaussian position probability density visible in Fig.~\ref{fig:c}.

Hence, the interference ripples have their origin in the phase interference of the parts with positive and negative energy.
As the phase moves with a superluminal average velocity $v_{\mathrm{ph}} = \lambda(\langle p\rangle)/\langle p\rangle$, also the interference pattern moves with superluminal speed.
Because it is just an interference pattern, it carries no information from one point in space to another.
Hence, despite the fact that local variations in the position probability density are, in principle, observable, the fact that these variations move at superluminal speed is no contradiction to the special theory of relativity.
A classical example for this effect is the apparent motion of Moir\'e patterns.
If two periodic patterns with slightly different periodicity lengths are slowly shifted with respect to each other, the pattern of the superposition (the Moir\'e pattern, which is locally observable) can move with almost arbitrary speed.
In fact the recently discussed phenomenon of superluminal tunneling also consists in the motion of an interference pattern of plane waves (thereby the transmitted wave packet changes its shape, but no information is being transmitted with superluminal speed).
A local perturbation of the interference pattern can only propagate with a velocity less than $c$.

\section{Analysis of Zitterbewegung}

The Zitterbewegung (that is, the shape of the white curve in the space-time diagram) is described by the expectation value (in the given initial state) of the operator
(see, for example, \cite{Thaller92})
\begin{equation}
x(t) = \rme{\rmi H_0 t} x \rme{-\rmi H_0  t} = 
x + v_{\mathrm{cl}}(p)\,t + Z(t).
\end{equation}
Here $v_{\mathrm{cl}}$ is the classical velocity operator \eqref{E: classical velocity operator},
and $Z(t)$ describes the oscillation,
\begin{equation}
Z(t) =  (2\rmi \,H_0)^{-1}\,(\rme{2\rmi H_0 t}{-}\id)\,\bigl(c{\sigma_1} {-} v_{\mathrm{cl}}(p)\bigr) .
\label{E: time evolution of x}
\end{equation}
Note that $c{\sigma_1}$ is the instantaneous velocity according to the Dirac equation,
because the time derivative of the position in the Heisenberg picture is just
\begin{equation}
\frac{d}{dt}\,x =  \rmi[H_0 , x ] = c\,\sigma_1.
\label{E: standard velocity operator}
\end{equation}
The initial wave packets \eqref{E: gaussian wave packet 1} and \eqref{E: gaussian wave packet 1} are eigenvectors of $\sigma_1$, belonging to the eigenvalue $+1$.
Hence the initial velocity of these wave packets is $+c$, hence in Fig.~\ref{fig:aworld} and Fig.~\ref{fig:bworld}, the slope of the white curve at $(x,t)=(0,0)$ is $45$ degrees.

In momentum space, $Z(t)$ is just a multiplication by a matrix-valued function of $p$.
The operator $Z(t)$ anticommutes with $H_0$.
This means that $Z(t)$ maps a state with positive energy onto a state with negative energy.
Suppose that $|\psi\rangle$ is, for example, a positive-energy state. Then $Z(t) | \psi\rangle$ has negative-energy and is thus orthogonal to $|\psi\rangle$.
Hence, the scalar product $\langle \psi | Z(t) | \psi\rangle$ must be zero.
The expectation value $\langle \psi | Z(t) | \psi\rangle$ can only be nonzero for wave packets having both positive- and negative-energy parts.

Wave packets located in different regions of momentum or position space are orthogonal.
The operator $Z(t)$ is a multiplication operator in momentum space which means that it does not change the localization properties of a wave packet in momentum space. Hence, Zitterbewegung can only be significant, if the function $\hat\psi_{\mathrm{pos}}(p,0)$ has a significant overlap with the function  $\hat\psi_{\mathrm{neg}}(p,0)$ in momentum space.

In position space, the operator $Z(t)$ is nonlocal, but it turns out, that it does not change the approximate localization of a wave packet all too much.
Hence, if a wave packet $\psi(x)$ is (approximately) located in some region $R$ of position space, then the function $Z(t)\psi(x)$ is approximately located in a neighborhood of that region.
As a consequence, Zitterbewegung is only significant as long as the parts with positive and negative energy are close to each other in position space.

Now, let us consider our examples.
Fig.~\ref{fig:amom} shows that the momentum distributions of the positive and negative energy parts of the initial wave packet \eqref{E: gaussian wave packet 1} have a significant overlap in momentum space.
This fact does not change with time, because the momentum (and hence the momentum distribution of the initial wave packet) is a conserved quantity according to the free Dirac equation.

Both parts of the initial wave packet have average momenta close to zero, corresponding to average (classical) velocities close to zero. Hence, the corresponding parts will remain close together also in position space. Hence, for this wave packet, Zitterbewegung is a sustained phenomenon.
(Actually, the amplitude of the Zitterbewegung vanishes very slowly, as $|t|\to \infty$.
A mathematical argument for the asymptotic decay of Zitterbewegung is given in \cite{Thaller92}.)

For the wave packet in Fig.~\ref{fig:b}, the momentum distributions of the positive and negative energy parts overlap completely (see Fig.~\ref{fig:bmom}).
Hence, we can indeed observe Zitterbewegung, but only as long as the two parts occupy approximately the same region in position space.
Because the two parts have opposite velocities, they quickly separate, and the amplitude of the Zitterbewegung decreases rapidly. 

For the wave packet in Fig.~\ref{fig:c}, the parts with positive and negative energy move together in position space, but there is no Zitterbewegung, because these parts are located in different regions of momentum space.

\section{Conclusion}

The interference effects occur only for those solutions of the Dirac equation that are composed of parts with positive and parts with negative energy. The origin for the ripples in the position distribution lies in the fact that the parts with positive and negative energy have opposite phase directions.

Because any square-integrable spinor-valued function (in particular, a Gaussian spinor) is, in general, a superposition of positive and negative energies, one is likely to run into these phenomena when one computes a numerical solution of the Dirac equation.

One may argue that superpositions of positive and negative energy states are not physically observable. A wave packet with only one sign of energies shows no Zitterbewegung and behaves reasonably. Indeed, in quantum field theory, one constructs a Hilbert space of many-particle electron states from the positive-energy solutions of the one-particle Dirac equation, and the positronic states are built from charge-conjugated negative-energy solutions. A problem with this approach is that the electronic and the positronic states cannot always be separated in the presence of external fields.

\end{document}